\documentclass[aip,apl,rsi,amsmath, amssymb,reprint]{revtex4-1}

\usepackage{natbib}
\usepackage{dcolumn}
\usepackage{graphicx,psfrag}
\usepackage{color}
\usepackage{amsfonts,amsmath,bm}
\usepackage{braket}
\usepackage{subfigure}

\begin{document}

\title{First-principles study on thermoelectric properties of half-Heusler compounds Co{\it M}Sb({\it M}$=$Sc, Ti, V, Cr, and Mn)}
\author{Susumu Minami}
\email{minami@cphys.s.kanazawa-u.ac.jp}
\affiliation{Graduate School of Natural Science and Technology, Kanazawa Univ., Kakuma, Kanazawa, 920-1192, Japan}
\author{Fumiyuki Ishii}
\email{fishii@mail.kanazawa-u.ac.jp}
\affiliation{Faculty of Mathematics and Physics, Institute of Science and Engineering, Kanazawa Univ., Kakuma, Kanazawa, 920-1192, Japan}
\author{Yo Pierre Mizuta}
\affiliation{Graduate School of Natural Science and Technology, Kanazawa Univ., Kakuma, Kanazawa, 920-1192, Japan}
\author{Mineo Saito}
\affiliation{Faculty of Mathematics and Physics, Institute of Science and Engineering, Kanazawa Univ., Kakuma, Kanazawa, 920-1192, Japan}
\date{\today}
\begin{abstract}
We have performed systematic density functional calculations and evaluated thermoelectric properties, Seebeck coefficient and anomalous Nernst coefficient of half-Heusler comounds Co{\it M}Sb({\it M}=Sc, Ti, V, Cr, and Mn).
 The carrier concentration dependence of Seebeck coefficients in nonmagnetic compounds are in good agreement with experimental values.
 We found that the half-metallic ferromagnetic CoMnSb show large anomalous Nernst effect originating from Berry curvature at the Brillouin zone boundary. 
 These results help to understanding for the mechanism of large anomalous Nernst coefficient and give us a clue to design high performance magnetic thermoelectric materials.
\end{abstract}


\maketitle

{\it Introduction.}
The Nernst effect induces a thermoelectric (TE) voltage under a magnetic field.
The direction of the TE voltage is perpendicular to the thermal gradient, which can be exploited to modularize TE generation devices.\cite{1882-0786-6-3-033003} 
Unlike the external magnetic field required for the conventional Nernst effect, the anomalous Nernst effect (ANE)\cite{ANE1,RevModPhys.82.1539,RevModPhys.82.1959} is induced by spontaneous magnetization.
This phenomenon has attracted attention as a new mechanism for TE generation systems.\cite{SAKURABA201629}

To realize widespread use of ANE-based TE generation devices, a large anomalous Nernst coefficient is needed.\cite{SAKURABA201629}
However, the reported anomalous Nernst coefficient $N$ ( $\le 1$ $\mu$V/K\cite{doi:10.1063/1.4922901,PhysRevLett.93.226601,PhysRevLett.99.086602,PhysRevLett.101.117208,PhysRevLett.107.216604,Mn3Sn}) is two orders of magnitude smaller than the Seebeck coefficient $S$ ($ \sim 10^2$ $\mu$V/K\cite{doi:10.1179/095066003225010182,HUANG2016107}) in typical TE materials.
 The ANE magnitude is mainly determined by the two factors in transport coefficient as follows: (i) the asymmetry of the anomalous Hall conductivity along the energy axis and (ii) the product of Seebeck coefficient and Hall angle ratio.\cite{doi:10.7566/JPSCP.3.017035,SkX} 
 These factors imply that large ANE could be found in magnetic materials with large anomalous Hall effect and/or large Seebeck effect.

Half-Heusler compounds are candidate materials for an ANE-based TE device.
Such compounds are known to be good TE materials with a large Seebeck effect originating from their narrow-gap semiconducting state with 18 valence electron counts per formula unit.\cite{GRAF20111}
For example, CoTiSb and NiTiSn show large $S$ of -320  and -250 $\mu$V/K at 300 K, respectively.\cite{TiCoSb_Seebeck,doi:10.1063/1.1318237}
Half-Heusler compounds are also well known as half-metallic ferromagnets with high Curie temperature.\cite{PALMSTROM2016371}
For example, CoMnSb and NiMnSb show ferromagnetism with Curie temperature of 490 and 730 K, respectively.\cite{CoMnSb_elcond}
Therefore, large ANE could be realized by tuning the electron or hole carrier of half-Heusler compounds.

In this study, we investigate the TE properties of half-Heusler compounds with the formula Co{\it M}Sb ({\it M} = Sc, Ti, V, Cr, and Mn). 
 We perform first-principles calculations for these compounds. 
 We calculate the TE properties based on the obtained electronic structures. 
 We estimate the carrier concentration dependence of the Seebeck coefficient by using semiclassical Boltzmann transport theory.
To clarify the ANE of ferromagnetic half-Heusler compounds, we focus on CoMnSb and estimate both the anomalous Nernst coefficient and the Seebeck coefficient.

\begin{figure}[htbp] \centering
  \includegraphics[width=\columnwidth]{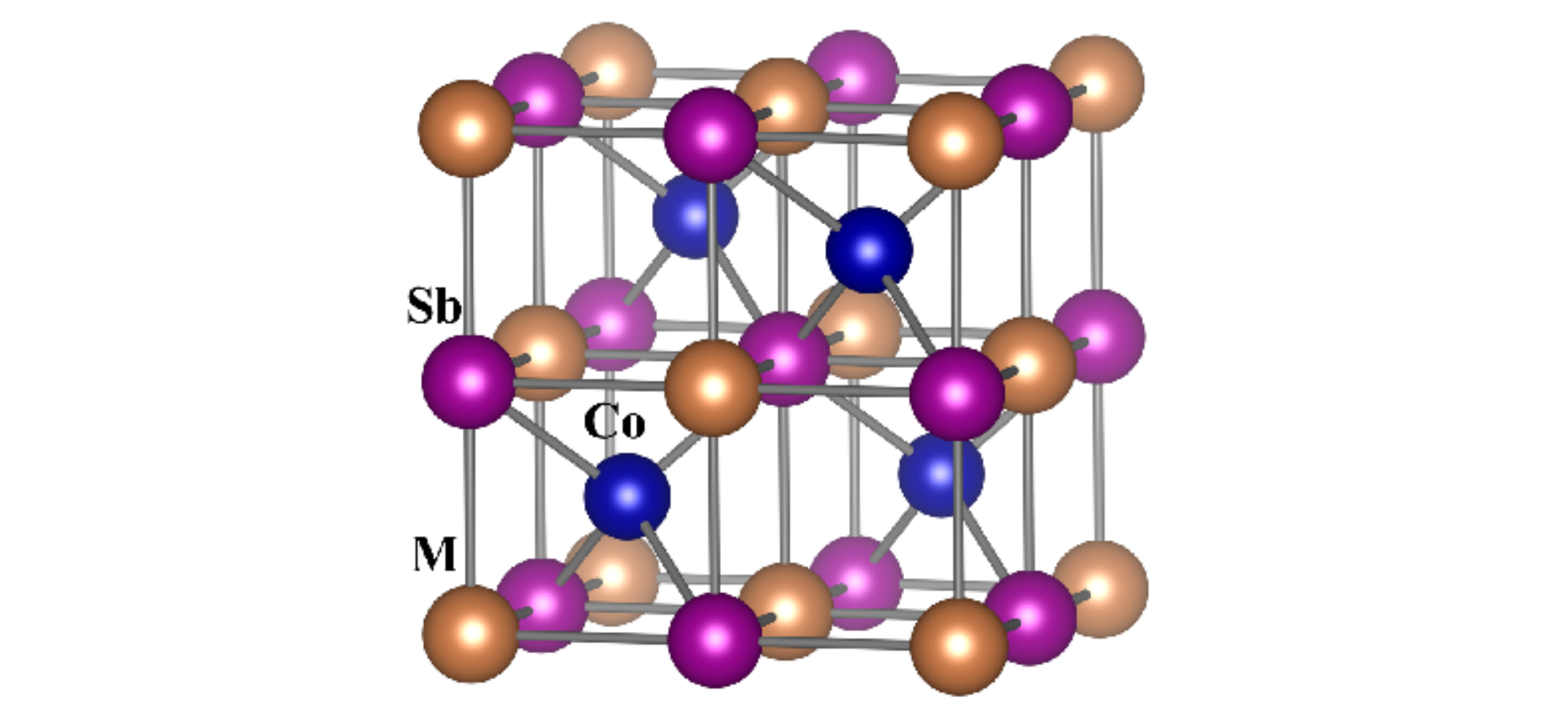}   
  \caption{\label{fig:structure}Schematic structure of the half-Heusler compounds Co{\it M}Sb.}
\end{figure}

{\it Computational Model.}
Figure \ref{fig:structure} shows the schematic structure of Co{\it M}Sb. 
Half-Heusler intermetallic compounds have a face-centered cubic crystal structure with chemical composition {\it XYZ } and space group $F\bar{4}3m$.  
The atomic sites in the unit cell {\it X}($\frac{1}{4}$,$\frac{1}{4}$,$\frac{1}{4}$), {\it Y}(0,0,0), and {\it Z}($\frac{1}{2}$, $\frac{1}{2}$, $\frac{1}{2}$) are occupied. 
The {\it X} atomic site is coordinated doubly tetrahedrally by four {\it Y} and four {\it Z}. Therefore, the {\it X} site is regarded as the unique site in the crystal structure. 
Table \ref{tab:lattice} shows the lattice constant, number of valence electron counts, and Curie temperature of Co{\it M}Sb. 
Because no experimental lattice constants have been reported for Sc and Cr, we used interpolated ones calculated based on those of {\it M} = Ti, V, and Mn. 
CoTiSb is a semiconductor with 18 valence electron counts and a narrow gap.
CoVSb and CoMnSb are ferromagnetic compounds with Curie temperatures of 58K and 490K, respectively.\cite{CoVSb_1972,CoMnSb_elcond}

\begin{table}[htb] \centering
\caption{\label{tab:lattice}Basic properties of Co{\it M}Sb. $a_{exp}$ and $a_{cal}$ are the experimental and theoretical lattice constant,  \ $n_{v}$ is the number of valence electrons per formula unit, $T_C$ is the Curie temperature. Our calculation were performed by using $a$ we estimated.}
  \begin{tabular}{cccccc} \hline \hline
    \it{M} &$a$ (\AA)& $a_{exp.} $(\AA) & $a_{calc}$ (\AA) & $n_v$ & $T_C$ (K) \\ \hline 
    Sc &6.06& - &6.09\cite{MgAgAs_calc1}& 17 &- \\ 
    Ti &5.88& 5.88\cite{TiCoSb_lc_ex} &5.88\cite{MgAgAs_calc2}& 18 & -  \\
     V &5.80& 5.80\cite{VCoSb_lc_ex} &5.81\cite{MgAgAs_calc2}& 19 & 58\cite{CoVSb_1972} \\
     Cr &5.79& - &5.79\cite{MgAgAs_calc1}& 20 & - \\
     Mn &5.87& 5.87\cite{MnCoSb_lc_ex} &5.82\cite{MgAgAs_calc2}& 21 & 490\cite{CoMnSb_elcond} \\ \hline \hline
  \end{tabular} 
  \end{table}

{\it Methods.}
We conducted first-principles calculations based on the non-collinear density functional theory\cite{NC1} (DFT) with OpenMX code.\cite{OpenMX}
DFT calculations are performed through the exchange-correlation functional of the generalized gradient approximation.\cite{PhysRevLett.77.3865}
We used norm-conserving pseudopotentials.\cite{Troullier_Efficient_1991}
The spin-orbit interaction (SOI) is included by using $j-$dependent pseudopotentials.\cite{PhysRevB.64.073106}
The wave functions are expanded by a linear combination of multiple pseudo-atomic orbitals.\cite{PhysRevB.67.155108}
The basis functions of each atoms are two $s$-, two $p$-, two $d$-,and one $f$- character numerical pseudo-atomic orbitals. 
The cutoff-energy for charge density is 250.0Ry. 
We use a $24\times24\times24$ uniform k-point mesh for self-consistent calculations. 
We construct maximally localized Wannier functions (MLWF) from DFT calculation results using Wannier90 code\cite{wannier90} and calculated the transport properties from the MLWF by using the semiclassical Boltzmann transport theory.\cite{ziman_1972}

The formulae for the TE coefficients can be derived from the linear response relation of charge current, $\bm j = \tilde{\sigma} \bm E + \tilde{\alpha}(-\nabla T)$, where $\bm E$ and $\nabla T$ are respectively the electric field and temperature gradient. 
By using the conductivity tensors $\tilde{\sigma} = [\sigma_{ij}]$ and $\tilde{\alpha} = [\alpha_{ij}]$, the Seebeck and anomalous Nernst coefficients are respectively expressed as\cite{doi:10.7566/JPSCP.3.017035,SkX}
\begin{eqnarray}
S \equiv S_{xx} \equiv \frac{E_x}{(\nabla T)_x} = \frac{S_{0} + \theta_{H}N_0}{1+\theta_{H}^2} \label{eq:Seebeck} \\
N \equiv S_{xy} \equiv \frac{E_x}{(\nabla T)_y} = \frac{N_{0} - \theta_{H}S_0}{1+\theta_{H}^2} \label{eq:Nernst} . 
\end{eqnarray}
Here we defined the conventional (pure) Seebeck, Hall angle ratio, and pure anomalous Nernst coefficient by using conductivity tensors for simplicity as follows: $S_{0} \equiv \alpha_{xx}/\sigma_{xx}$, $\theta_{H} \equiv \sigma_{xy}/\sigma_{xx}$, $N_{0} \equiv \alpha_{xy}/\sigma_{xx}$, respectively. 
The longitudinal conductivity tensor is calculated as $\sigma_{xx} $=$ e^2 \tau \sum_n \int d\bm k v_{x}^n(\bm k)^2 \left( -\frac{\partial f}{\partial \varepsilon_{n\bm k}} \right)$, and
the transverse conductivity tensor is calculated as $\sigma_{xy} $=$ -\frac{e^2}{\hbar}\sum_n \int d\bm k \Omega_z^n(\bm k) f(\varepsilon_{n\bm k})$.
Both the longitudinal and the transverse TE conductivity tensor are calculated as $\alpha_{ij} $=$ \frac{1}{e}\int d\varepsilon \sigma_{ij}(\varepsilon)|_{T=0} \frac{\varepsilon - \mu}{T}\left( -\frac{\partial f}{\partial \varepsilon} \right)$.
In the above formula, $\tau, f, v_x^n, \varepsilon_{n\bm k}, \Omega_{z}^n$, and $\mu$ denote the relaxation time (assumed to take constant value $\tau$), the Fermi-Dirac distribution function, group velocity of electrons, energy and $\bm k$-space Berry curvature, and chemical potential, respectively. 
The Berry curvature is determined by $\bm \Omega^n(\bm k) \equiv i\bra{\bm \nabla_{\bm k} u_{n\bm k}}\times \ket{\bm \nabla_{\bm k} u_{n\bm k}}$. 
The subscript index $n$ is the band index. 
Note that the conventional Seebeck coefficient $S_0$ is calculated using $\theta_{H} = 0$ and $N_{0} = 0$ in Eq.(\ref{eq:Seebeck}). 
We focus on the intrinsic contribution of $\sigma_{xy}$ and neglect the extrinsic one caused by impurities or defects. 
Here, in order to discuss the chemical potential ($\mu$) dependence of TE coefficients, we introduce Mott's formula,  $S_0 \simeq \alpha^{(1)}_{xx}/\sigma_{xx}$, $N_0 \simeq \alpha^{(1)}_{xy}/\sigma_{xx}$, where $\alpha_{ij}^{(1)}$, defined as $\alpha_{ij}^{(1)}(\mu) \equiv - \frac{\pi^2}{3}\frac{k^2_BT}{e} \frac{\partial \sigma_{ij}(\varepsilon)}{\partial \varepsilon} |_{\varepsilon=\mu}$ with Boltzmann's constant $k_B$, is the low $T$ approximation to $\alpha_{ij}$.

\begin{figure}[htb]
  \centering
  \includegraphics[width=\columnwidth]{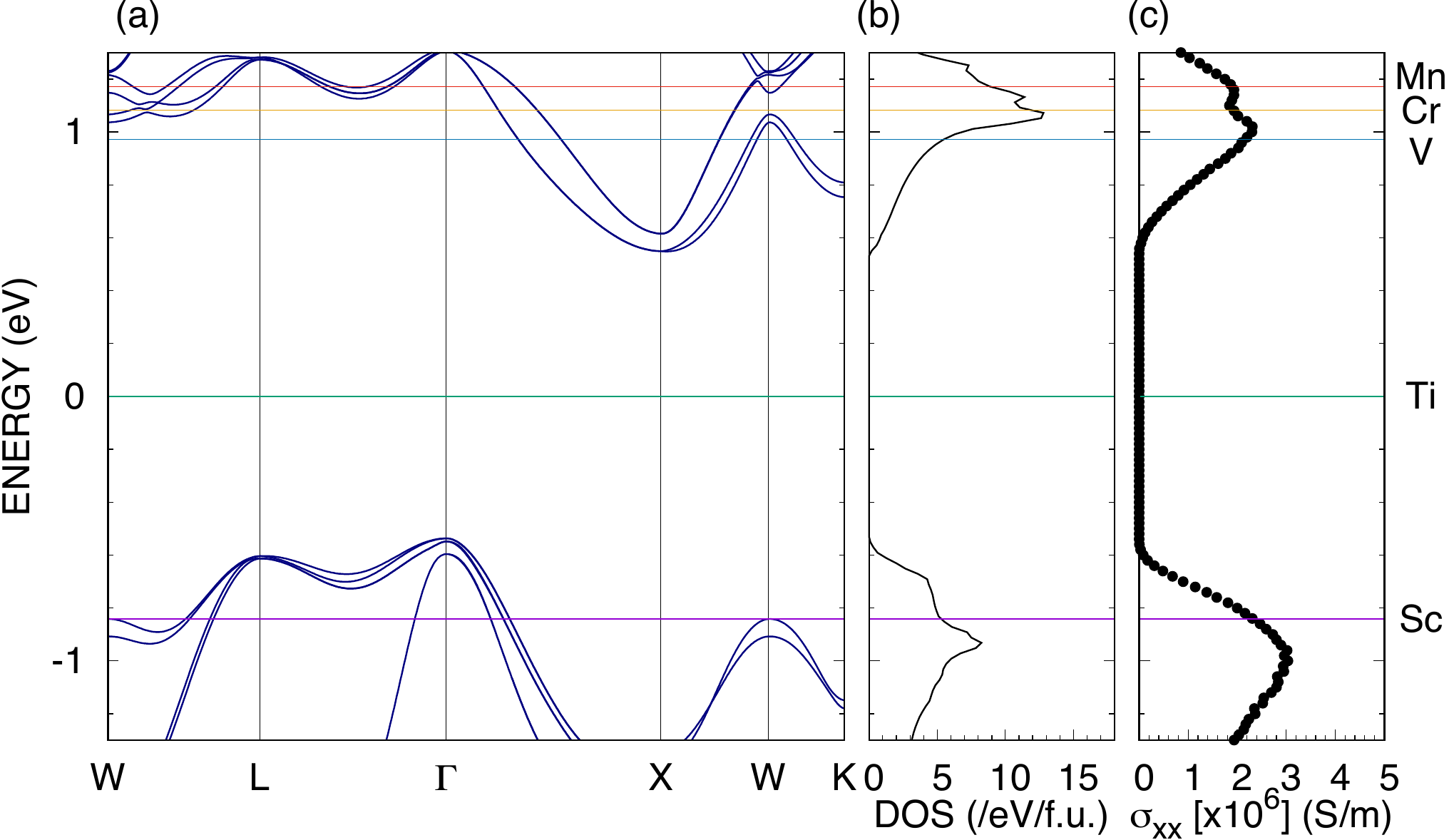}
  \caption{\label{fig:TiCoSb_band}Band structure (a), density of states (b), and electrical conductivity at 0K with relaxation time $\tau = 10$ fs(c) of CoTiSb. The origin of the energy is taken to Fermi energy for CoTiSb. The color lines show the Fermi energy of Co{\it M}Sb according to the rigid band approximation.}
\end{figure}

\begin{figure}[htb]
   \includegraphics[width=.9\columnwidth]{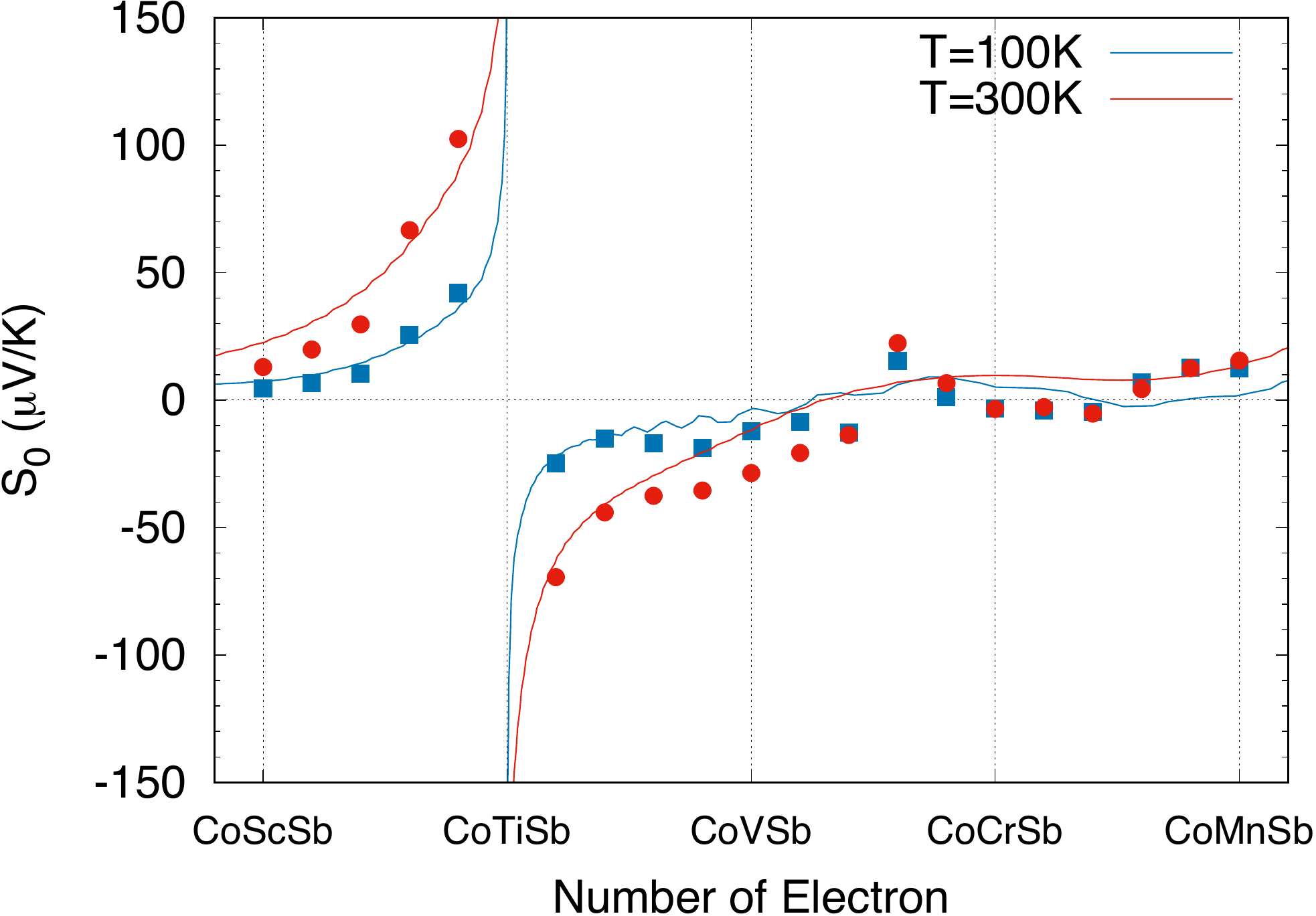}
  \caption{\label{fig:Seebeck_NM}Carrier concentration dependence of Seebeck coefficient in Co{\it M}Sb at 100K and 300K. The solid lines show the Seebeck coefficient according to rigid band approximation. Squares and circles show the Seebeck coefficient calculated by using electronic structure with self-consistent field calculation.}
\end{figure}

{\it Nonmagnetic phase.}
Because $N_0$ and $\theta_H$ are zero for nonmagnetic materials (CoScSb and CoTiSb) and the nonmagnetic (paramagnetic) phase of CoVSb, CoCrSb, and CoMnSb, we first calculate the conventional (pure) Seebeck coefficient $S_0$ at first.

Figure \ref{fig:TiCoSb_band} shows the electronic structure, density of states (DOS), and electrical conductivity at 0K of CoTiSb.
The calculated band gap for 1.06 eV is in good agreement with the experimental band gap of 0.95 eV.\cite{TICoSb_exp_1998} 
The Fermi energies ($E_{\rm F}$) are calculated by the rigid band approximation (RBA) for other half-Heusler compounds, namely, CoScSb, CoVSb, CoCrSb, and CoMnSb.

Figure \ref{fig:Seebeck_NM} shows carrier concentration dependence of $S_0$. 
$S_0$ is estimated by two approaches: (i) RBA in CoTiSb and (ii) self-consistent field (SCF) calculation by hole- or electron-doping around each pristine Co{\it M}Sb. 
The trends of the carrier concentration dependence of $S_0$ are broadly consistent between RBA and SCF.

 Positive (negative) $S_0$ values are observed on the left (right) side of CoTiSb in Fig. \ref{fig:Seebeck_NM}, which can be attributed to as hole- (electron-) doped semiconductor, respectively. 
 The $S_0$ values of CoTi$_{0.95}$Sc$_{0.05}$Sb and CoTi$_{9.95}$V$_{0.05}$Sb at 300K as calculated by RBA is 138$\mu$V/K and -125$\mu$V/K, respectively; these are in good agreement with the experimental values of 178 $\mu$V/K and -163$\mu$V/K.\cite{PhysRevB.86.045116}
 On the other hand, the Seebeck coefficient calculated by SCF of CoVSb is around three times larger than the calculated RBA, which is much closer to the experimental value\cite{CoVSb_exp_2001} of -45 $\mu$V/K.

The trend of the carrier concentration dependence of $S_0$ can be roughly interpreted using Mott's formula.
Because $\sigma_{xx}$ at $E_{\rm F}$ of CoScSb, CoCrSb, and CoMnSb has negative slope in Fig. \ref{fig:TiCoSb_band}(c), $\alpha_{xx}^{(1)}$ is positive. 
On the other hand, because $\sigma_{xx}$ at $E_{\rm F}$ of CoVSb  has positive slope, $\alpha_{xx}^{(1)}(E_{\rm F})$ is negative. 
For {\it M}= Cr, the sign of $S_0$ between RBA and SCF is different, implying that RBA using the band structure of CoTiSb is not appropriate for CoCrSb.

\begin{figure}[htb] \centering
  \includegraphics[width=\columnwidth]{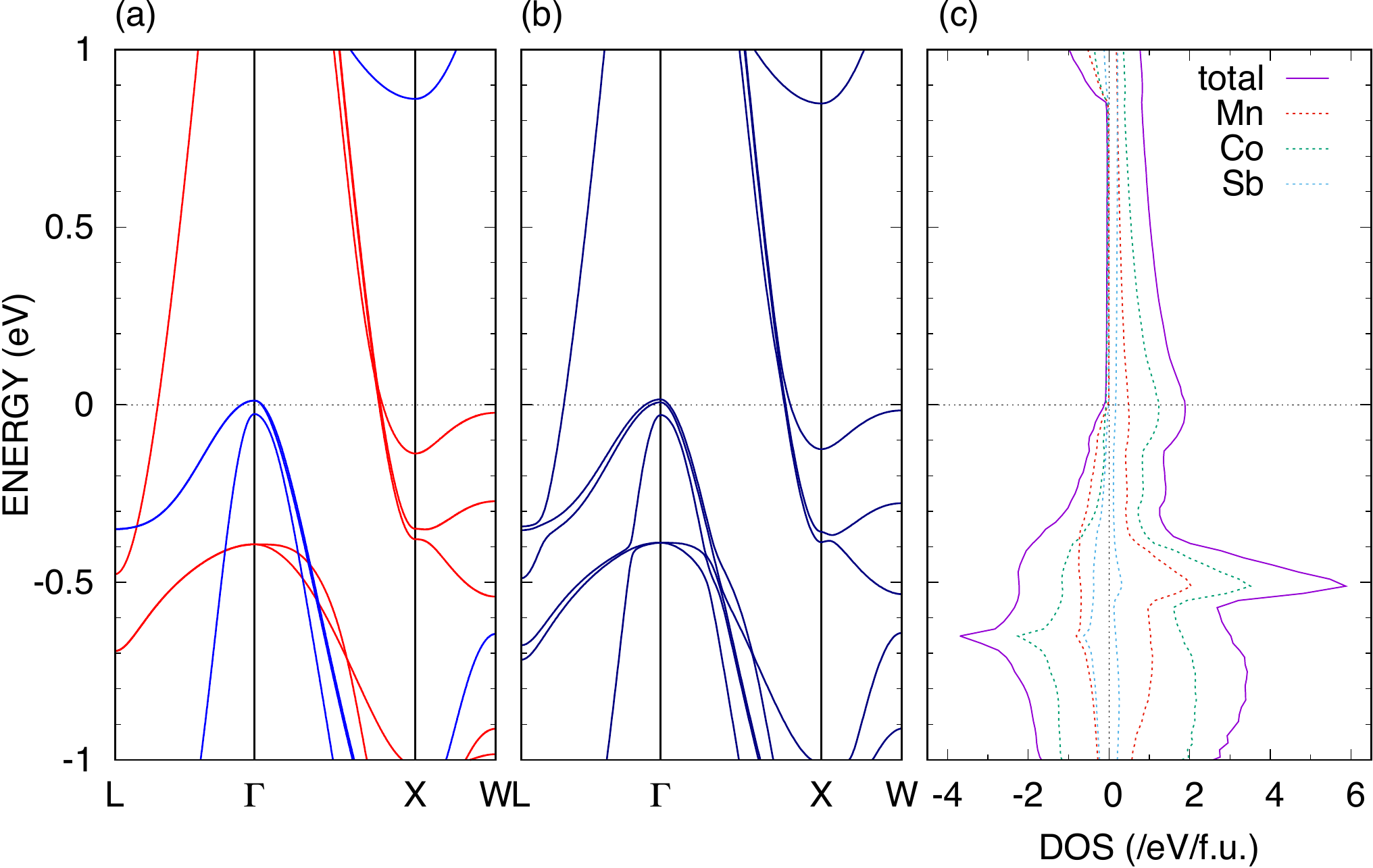}
  \caption{\label{fig:CoMnSb_band-dos} Band structure without SOI (a), with SOI (b), total and projected density of states(c) for CoMnSb. Blue and red lines show the majority and minority spin, respectively.  The Fermi energy is set to 0 eV.}
\end{figure}

\begin{table}[htb] \centering
\caption{\label{tab:TEcoefficient}Each component of calculated thermoelectric coefficients ($\mu$V/K), Hall angle ratio, and evaluated relaxation time (fs) for CoMnSb at Fermi energy($\mu = 0$).}
  \begin{tabular}{ccccccc} \hline \hline
     Temperature(K)&$S_{0}$& $N_{0}$ &  $\theta_{H}[\times 10^{-2}]$ & $S$ &$N$ & $\tau$ \\ \hline 
     100 & - 5.80 &  -0.11 & -0.42 & -5.79 & -0.13 & 7.0\\
     300 & -16.00  &  -0.85 & -1.02 & -15.99 & -1.02 & 2.9 \\ \hline \hline
  \end{tabular} 
  \end{table}

\begin{figure}[htb] \centering
  \includegraphics[width=\columnwidth]{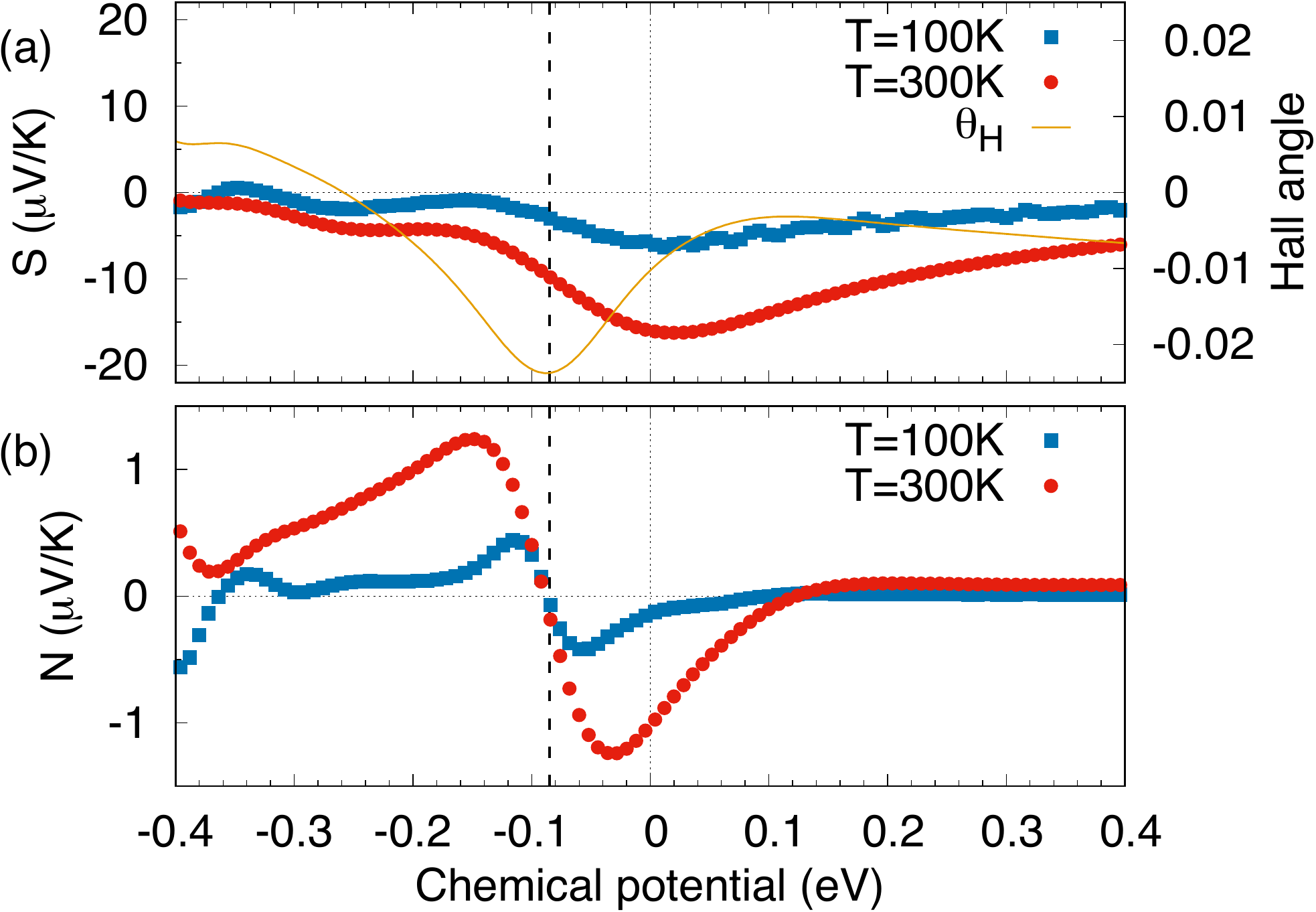}
          \caption{\label{fig:CoMnSb_ane}Thermoelectric properties for CoMnSb at 100K and 300K. Each panel show Seebeck coefficient and Hall angle ratio(300K) (a), anomalous Nernst coefficient (b).}
\end{figure}

\begin{figure}[htb] \centering
  \includegraphics[width=\columnwidth]{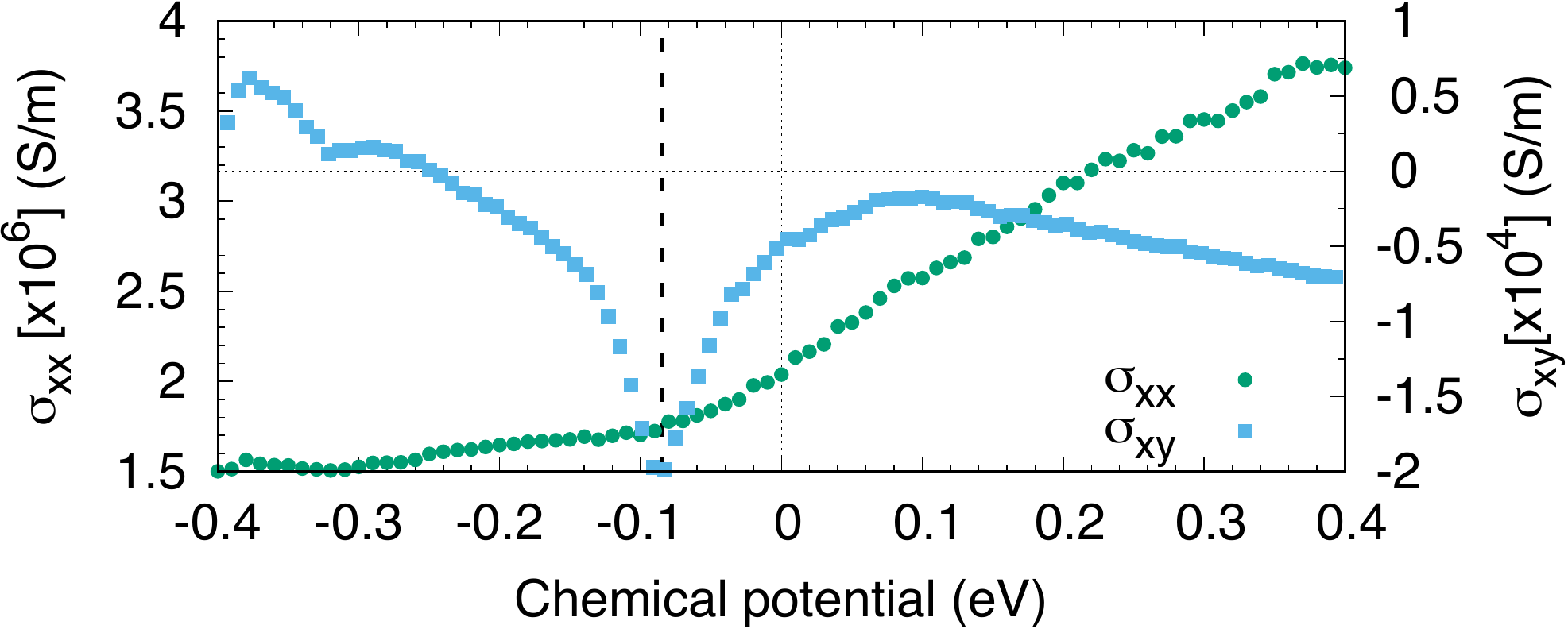}
          \caption{\label{fig:CoMnSb_elcond} Chemical potential dependence of electrical conductivities $\sigma_{xx}$(S/m) with relaxation time $\tau=10$ fs and $\sigma_{xy}$(S/m) at 0K .}
\end{figure}

{\it Ferromagnetic phase in CoMnSb.}
We investigate CoMnSb because of its applicability for an ANE-based TE module that can operate below room temperature.
Figure \ref{fig:CoMnSb_band-dos} show the band structure and DOS for CoMnSb .  
At the Fermi level, the majority (minority) spin shows an electron- (hole-) like band structure around the X ($\Gamma$) point in Figs. \ref{fig:CoMnSb_band-dos}(a) and (b).
A comparison of the DOS at $E_{\rm F}$ of the nonmagnetic and ferromagnetic phases in Fig. \ref{fig:TiCoSb_band}(b) and Fig. \ref{fig:CoMnSb_band-dos}(c) indicates that the ferromagnetic phase has smaller $D(E_{\rm F}) \sim 2$ /eV/f.u. than $D(E_{\rm F}) \sim 9$/eV/f.u. in the nonmagnetic phase.  
This change in $D(E_{\rm F})$ can be attributed to ferromagnetic phase transition induced by Stoner instability. 
 The total magnetic moment is  $\sim 3\mu_B$/f.u..
 Moreover, the atomic magnetic moments of Co, Mn, and Sb are obtained -0.3, 3.6, and -0.3 $\mu_B$/f.u., respectively.

Table \ref{tab:TEcoefficient} shows the TE properties of CoMnSb at $E_{\rm F}$.
The relaxation times $\tau$ are estimated as $\tau=\tau_{0}(\rho_{calc}/\rho_{exp})$, where $\rho_{calc}=1/\sigma_{xx}(\tau_0)$ is the calculated electrical resistivity and $\rho_{exp}$ is the experimental one reported in Ref. \onlinecite{CoMnSb_elcond}.
For CoMnSb, we estimated as $\tau{\rm (100K)}=7.0$ fs and $\tau{\rm (300K)}=2.9$ fs. 
We found unique TE properties in  the ferromagnetic phase.
Furthermore, the ferromagnetic phase shows negative $S$ whereas the nonmagnetic phase shows positive S (Fig. \ref{fig:Seebeck_NM}).
A large anomalous Nernst coefficient ($N$) that reaches  -1.02 $\mu$V/K at 300K was also found; this value is fairly large compared to that of reported ferromagnetic metals,  for example,  $\sim 0.6\mu$V/K for L1$_0$-ordered FePt thin film\cite{doi:10.1063/1.4922901}.
The main component of $N$ (referring to Eq. (\ref{eq:Nernst})) is the pure anomalous Nernst coefficient ($N_0$). 
The contributions of two components $N_0$ and $S_0\theta_H$ are  $\sim 80\%$ and $\sim20\%$, respectively.

To understand the unique TE properties in the ferromagnetic phase of CoMnSb, we show the chemical potential dependence of the TE coefficient at 100 and 300 K  and the electrical conductivity at 0 K in Figs. \ref{fig:CoMnSb_ane} and \ref{fig:CoMnSb_elcond}, respectively.
First, we discuss the Seebeck coefficient ($S$).
The negative sign of $S$ at $E_{\rm F}$ can be understood from Mott's formula:  $\alpha_{xx}^{(1)}(E_{\rm F})$ is negative because $\sigma_{xx}(E_{\rm F})$ has positive slope, as shown in Fig. \ref{fig:CoMnSb_elcond}.
The peak of $S$ in Fig. \ref{fig:CoMnSb_ane}(a) is around $E_{\rm F}$, and the maxima is $\sim 15 \mu$V/K at 300 K, which is not so large compared with that of typical TE materials. 
Next, we discuss the anomalous Nernst coefficient ($N$).
$N$ shows two peaks in Fig. \ref{fig:CoMnSb_ane}(b) at the lower- and higher-energy sides of $\mu_P \equiv -85$ meV. 
The chemical potential of $\mu_{P}$ is indicated by the vertical dotted line in Figs. \ref{fig:CoMnSb_ane} and \ref{fig:CoMnSb_elcond}.
For $N$, the two peaks show opposite signs owing  to the almost-even functional form $\sigma_{xy}(\varepsilon-\mu_P)\simeq \sigma_{xy}(-\varepsilon + \mu_P)$, leading to the almost-odd functional form of its first derivative, to which $\alpha^{(1)}_{xy}$ is proportional as shown in Mott's formula.

\begin{figure}[htb] 
  \begin{minipage}{\columnwidth}
            \includegraphics[width=\columnwidth]{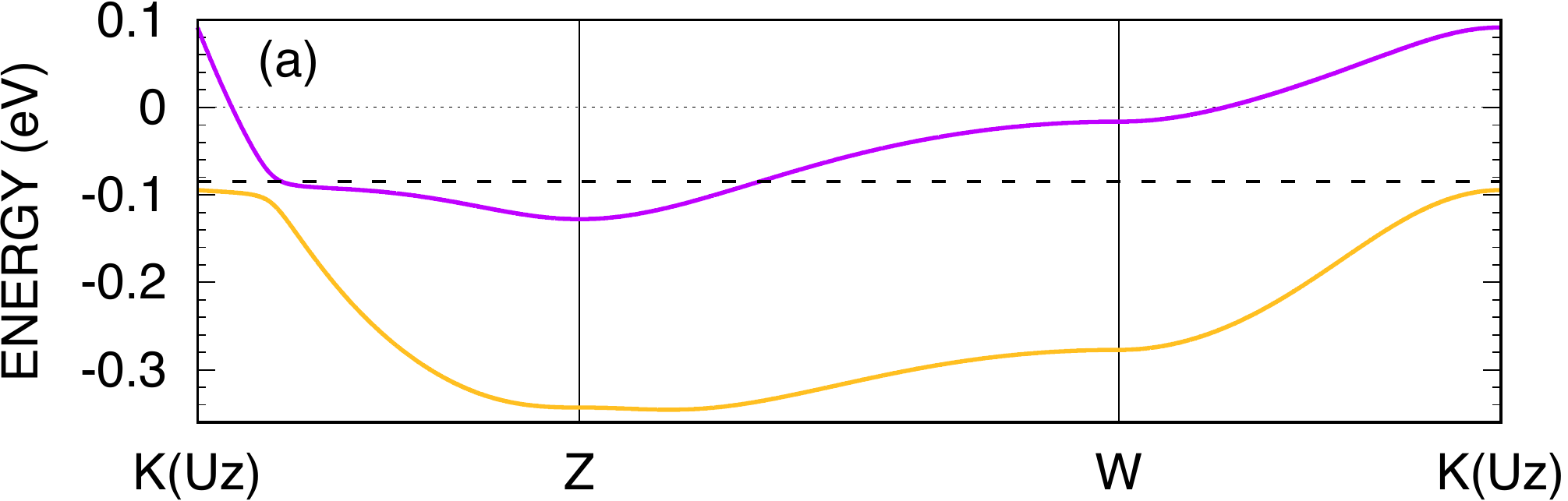}%
  \end{minipage} \\
  \begin{minipage}{\columnwidth}
            \includegraphics[width=.7\columnwidth]{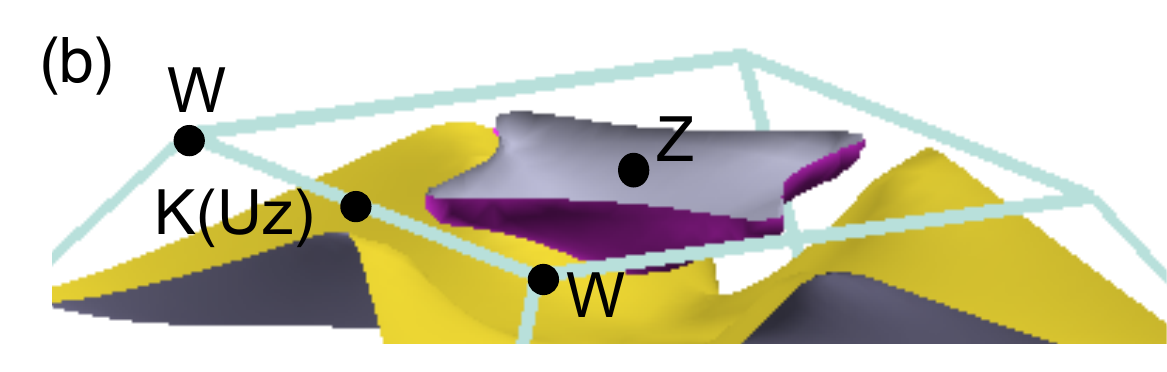}%
  \end{minipage} \\  
  \begin{minipage}{\columnwidth}
    \includegraphics[width=.9\columnwidth]{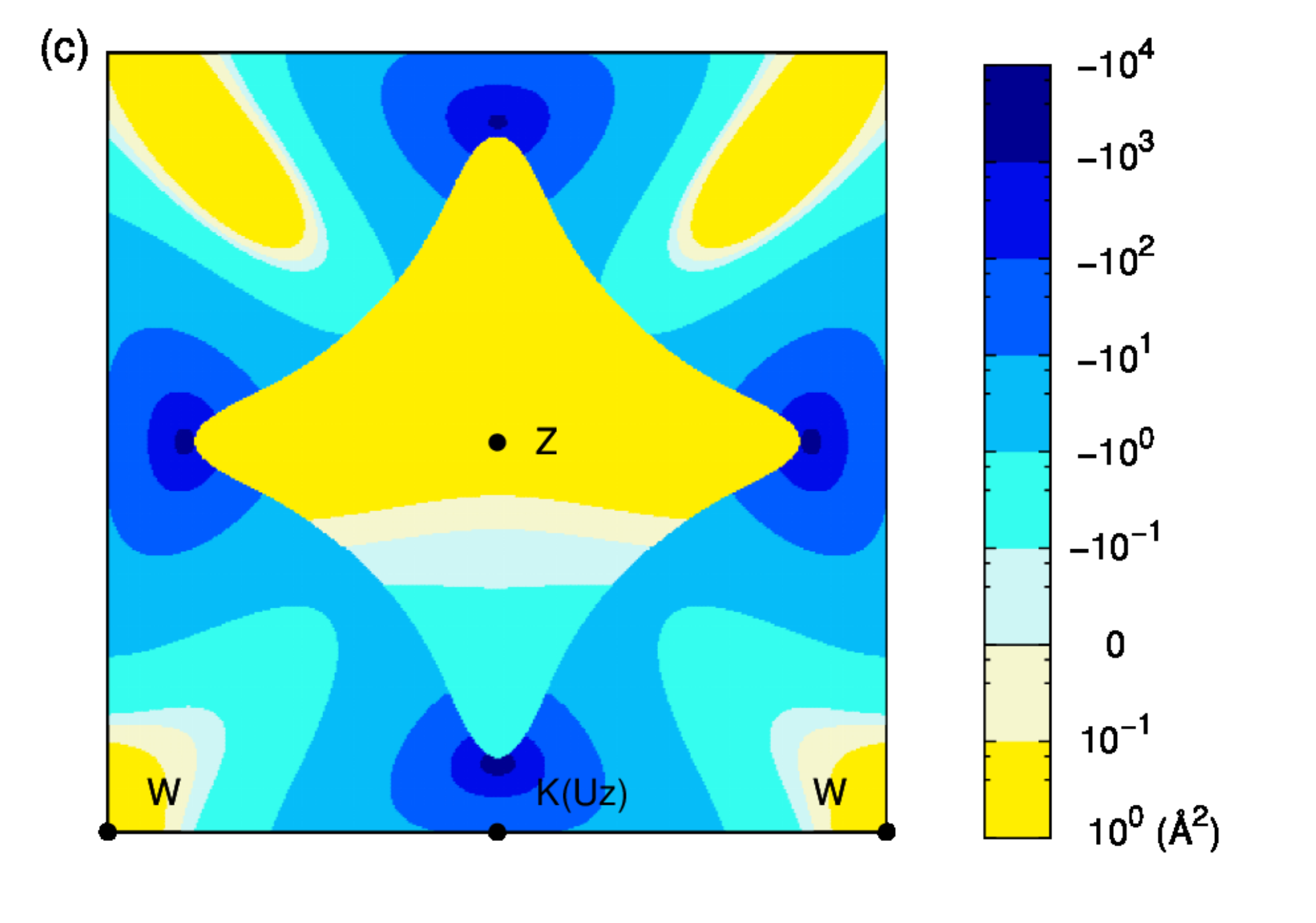} %
  \end{minipage}   
  \caption{\label{fig:CoMnSb_berry}Band structure with SOI (a), iso-energy surface for $\varepsilon=-85$ meV (b), and Sum of Berry curvature over occupied states $\overline{\Omega}_k \equiv \sum_n -\Omega_z^n(\bm k) f_n(\varepsilon) $ in the $k_z= \frac{2\pi}{a}$ plane (c) for CoMnSb.}
\end{figure}

To clarify the origin of the peak of $\sigma_{xy}$, we focus on the iso-energy surface at $\mu_P$ because it is well known that a large Berry curvature appear results from the crossing band at the Fermi level.\cite{PhysRevB.85.012405,1367-2630-15-3-033014}
Figure \ref{fig:CoMnSb_berry}(a) shows the band structure around $\mu_P$ corresponding to the horizontal dotted line. 
The symmetry point denoted as Z($0,0,\frac{2\pi}{a}$) is equivalent to X($0,\frac{2\pi}{a},0$) in the absence of magnetic ordering; however, some distinctively points appear because of the magnetic ordering. 
For example, we confirmed that the energy is shifted by $\sim 20$ meV between the iso-energy surface on the Z($0,0,\frac{2\pi}{a}$)-Uz($\frac{\pi}{2a},\frac{\pi}{2a},\frac{2\pi}{a}$) and X($0,\frac{2\pi}{a},0$)-Ux($\frac{\pi}{2a},\frac{2\pi}{a},\frac{\pi}{2a}$) lines.

The iso-energy surface and summed Berry curvature around Z($0,0,\frac{2\pi}{a}$) are shown in Fig. \ref{fig:CoMnSb_berry}(b) and Fig. \ref{fig:CoMnSb_berry}(c), respectively.
The asymmetry of the Berry curvature in Fig. \ref{fig:CoMnSb_berry}(c) is induced by the magnetic moment that is slightly canted from the z-axis.
Largely negative summed Berry curvature ($\overline{\Omega}_k \sim -10^4$) appears near Uz($\frac{\pi}{2a},\frac{2\pi}{a},\frac{\pi}{2a}$). 
This peak shows that the summed Berry curvature changes discontinuously at these boundaries, indicating that the crossing band (corresponding to purple one in Fig. \ref{fig:CoMnSb_berry}(a)) has large positive Berry curvature. 
Furthermore, it can be predicted that another crossing band (corresponding to yellow one in Fig. \ref{fig:CoMnSb_berry}(a)) also has large negative Berry curvature. 
It is obvious that the peak of $N$ and $\sigma_{xy}$ result from the change in the Berry curvature on the Z($0,0,\frac{2\pi}{a}$)-Uz($\frac{\pi}{2a},\frac{2\pi}{a},\frac{\pi}{2a}$) line.

{\it Conclusion.}
In summary,  systematic DFT calculations are used to determine the carrier concentration dependence of the TE properties in Co{\it M}Sb ({\it M}=Sc, Ti, V, Cr, and Mn). 
In the nonmagnetic phase, the calculated Seebeck coefficient of  CoTi$_{0.95}${\it M}$_{0.05}$Sb ({\it M}=Sc,V) shows good agreement with the experimental data. 
In the ferromagnetic phase, we focus on half-metailc CoMnSb because of its high $T_{\rm C}$.
The Seebeck coefficient shows opposite sign to the nonmagnetic phase.
Furthermore, the large anomalous Nernst coefficient ($N$) reaches -1.02$\mu$V/K.
We conclude that the peaks of $N$ originate from the large Berry curvature  on the Z-Uz line.  
These results should help in understanding for the mechanism of large ANE in half-Heusler compounds.

\clearpage
\bibliography{./ref}

\end{document}